\documentclass[%
 reprint,
 amsmath,amssymb,
 aps,
pra,
onecolumn]{revtex4-1}
\usepackage{graphicx}
\begin{document}

\title{Unconventional superconductivity after the BCS paradigm and empirical rules for the exploration of high temperature superconductors}

\author{Hai-Hu Wen}

\address{Center for Superconducting Physics and Materials,
National Laboratory of Solid State Microstructures and Department
of Physics, Nanjing University, Nanjing 210093, China}

\email{hhwen@nju.edu.cn}

\begin{abstract}
Superconducting state is achieved through quantum condensation of
Cooper pairs which are new types of charge carriers other than single electrons in normal metals. The theory established by Bardeen-Cooper-Schrieffer (BCS) in 1957 can successfully explain the phenomenon of superconductivity in many
single-element and alloy superconductors. Within the BCS scheme, the Cooper pairs are formed by exchanging the virtual vibrations of lattice (phonons) between two electrons with opposite momentum near the Fermi surface. The BCS theory has dominated the field of
superconductivity over 64 years. Many superconductors discovered in
past four decades, such as the heavy Fermion superconductors, cuprates, iron pnictide/chalcogenide
and nickelates seem, however, to strongly violate the BCS
picture. The most important issue is that, perhaps the BCS picture based on electron-phonon coupling are the special case for superconductivity, there are a lot of other reasons or routes for the Cooper pairing and superconductivity. In this short overview paper, we will summarize part of these progresses and try to guide readers to some new possible schemes of superconductivity after the BCS
paradigm. We also propose several empirical rules for the exploration of high-temperature unconventional superconductors.
\end{abstract}

\maketitle

\section{Introduction}
Superconductivity is a very interesting quantum state which exhibits zero resistivity and expulsion of magnetic field (Meissner effect). It has been well understood that this fascinating phenomenon is induced by the condensation of electron pairs (termed as Cooper pairs) at a finite temperature. Up to date, the superconducting state can only be achieved with these electronic
Cooper pairs although some other novel pictures concerning other
type of charged bosons have been proposed. Since its first discovery
in April 1911 by Kamerling Onnes et al., the field of
superconductivity has experienced glorious
developments and achievements in past 111 years. Owing to the deep and interesting
physics involved by the phenomenon of superconductivity, and the
potential of large scale applications, the field has
been pushed forward step by step, most of time by the discovery of
new superconductors. We can expect more interesting phenomena and materials, which will eventually lead to the
revolution, both on the fundamental understanding of condensed matter
physics and bring in the dreamful industrial applications of
superconductors.

\begin{figure}

\vspace{-1.0em} %
\centering
\setlength{\abovecaptionskip}{0.cm}
\includegraphics[width=12cm]{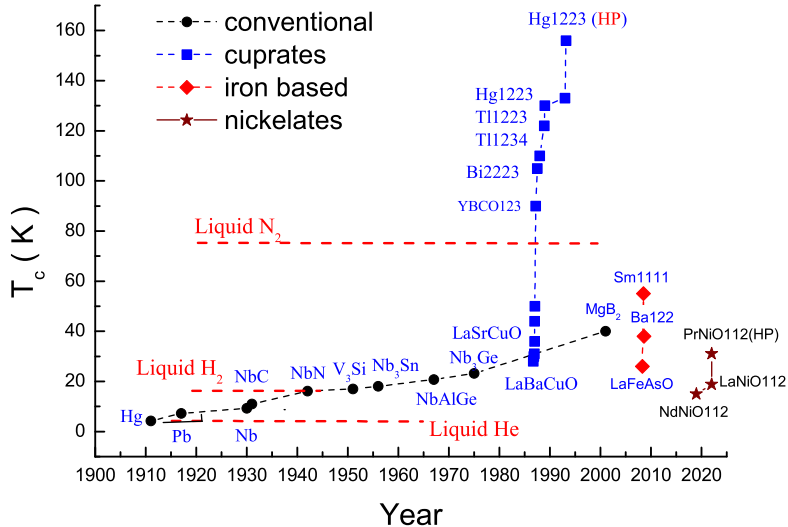}
\caption{\label{Figure1}Correlation between the superconducting transition
temperatures and the year for them to be found.
The three horizontal red dashed lines represent the boiling
temperatures of liquid helium, hydrogen, and nitrogen. The circles, squares, diamonds, stars represent the conventional, cuprate, iron-based and nickelate superconductors, respectively.}\label{f1}
\end{figure}

The development of the field of superconductivity is however not
straightforward, it is always mixed with some excitement and
frustrations. In Fig.\ref{f1}, we present the time dependence of the
superconducting transition temperatures T$_c$ of some representative
superconducting systems when they were discovered. One can see that,
the value of T$_c$ ramps up with a small slope before October 1986.
After the tremendous efforts in 75 years, the T$_c$ finally reached
about 23.2 K in $Nb_3Ge$ system. Most superconductors discovered before this time were single element metals or alloys. The basic reason
for superconductivity, called as the superconducting mechanism, was
unraveled in 1957 by Bardeen, Cooper and Schrieffer and named as the
BCS theory\cite{BCS}. In this short review, we omit the description about the BCS paradigm. A short review can be found in our earlier review paper\cite{WenHHJPS2014}. Since the field is rapidly moving forward, the understanding on the so-called
unconventional superconductivity have not come to the mature status
yet. Thus this paper serves only as a brief introduction and may be taken as a
supplementary to other beautiful reviews in this fascinating field. Nowadays, it is known that either in conventional or unconventional superconductors, we have an energy gap that protects the superconducting condensate from breaking Cooper pairs or exciting so-called quasiparticles. In the BCS theory, the gap is given by

\begin{equation}
\Delta_{s}= 2\hbar\omega_D exp^{-1/N(0)V}.
\label{e4}
\end{equation}

Here $\hbar \omega_{D}$ = $k_B\Theta_D$ is the Debye energy; N(0) is the density of states (DOS) near the Fermi energy; $V$ is the interaction between two electrons involved in the pair-scattering process. In the weak coupling case, it is predicted that $\Delta_s=1.75
k_BT_c$. Therefore the superconducting state can be achieved at a
finite temperature. One can see from the above description that, one of the basic conditions in the establishment of the BCS scheme is that $\Delta_s<<\hbar\omega_D<<E_F$. Once we have a system with $\Delta_s$ comparable with the Fermi energy $E_F$, this basic requirement is thus lost and we run into a new scheme of superconductivity. For superconductors with conventional superconducting pairing mechanism, the one with the highest transition temperature at ambient pressure is $MgB_2$ which was discovered in 2001\cite{MgB2}. In some pressurized hydrides\cite{Hydride}, it was reported that the superconducting transition temperature rises to about 200K or beyond, but the pairing mechanism can still be described by the phonon mediated BCS picture.

\section{Unconventional superconductivity in cuprates and iron pnictides/chacogenides}
In past four decades, many new superconductors were discovered.
These systems exhibit some features which are clearly at odds with
the BCS theory. These superconductors include the heavy
Fermion\cite{HeavyFermion}, organic\cite{Organic}, cuprates,
iron based superconductors and nickelates. Among them, the cuprate is a typical one
which was firstly discovered by K. A. M\"{u}ller and J. G.
Bednorz\cite{Cuprate}. They found that the superconductivity
occurred at about 35 K in a typical copper oxide
$La_{2-x}Ba_xCuO_4$. This discovery ignited an explosion in the
field of superconductivity. In later 7-8 years, many other new
superconducting systems were discovered, for example the
$YBa_2Cu_3O_{7-\delta}$ ($T_c\approx 90 K$)\cite{ChuCW,ZhaoZX}, the
bismuth family $Bi_2Sr_2CaCu_2O_8$ ($T_c\approx90 K$) and
$Bi_2Sr_2Ca_2Cu_3O_{10}$ ($T_c\approx125 K$)\cite{BiSystem}, the
thallium family\cite{TlSystem} and the mercury family
\cite{HgSystem} etc. So far the highest superconducting transition
temperature in cuprates is T$_c$ = 164 K, which was discovered in the
$HgBa_2Ca_2Cu_3O_{8+\delta}$ under a high pressure of 45
GPa.\cite{164K}.

\begin{figure}
\vspace{-2.0em} %
\centering
\setlength{\abovecaptionskip}{-3.0em}
\includegraphics[width=10cm]{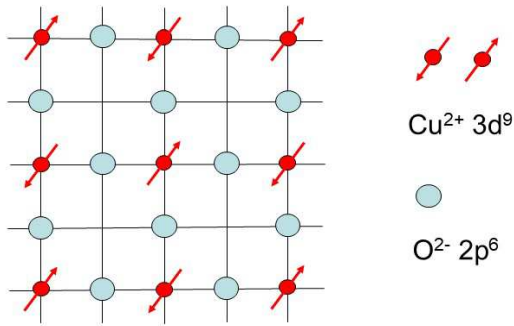}
\caption{Sketch of the $CuO_2$ planes of the cuprate
superconductors. The small filled circles with arrows represent the
Cu ions with the $3d^9$ as the outermost orbital of electrons. The
large circles represent the oxygen ions with the 2p$^6$ outermost
orbitals. } \label{f2}
\end{figure}

\begin{figure}
\vspace{0.0em} %
\centering
\setlength{\abovecaptionskip}{0.0em}
\includegraphics[width=12cm]{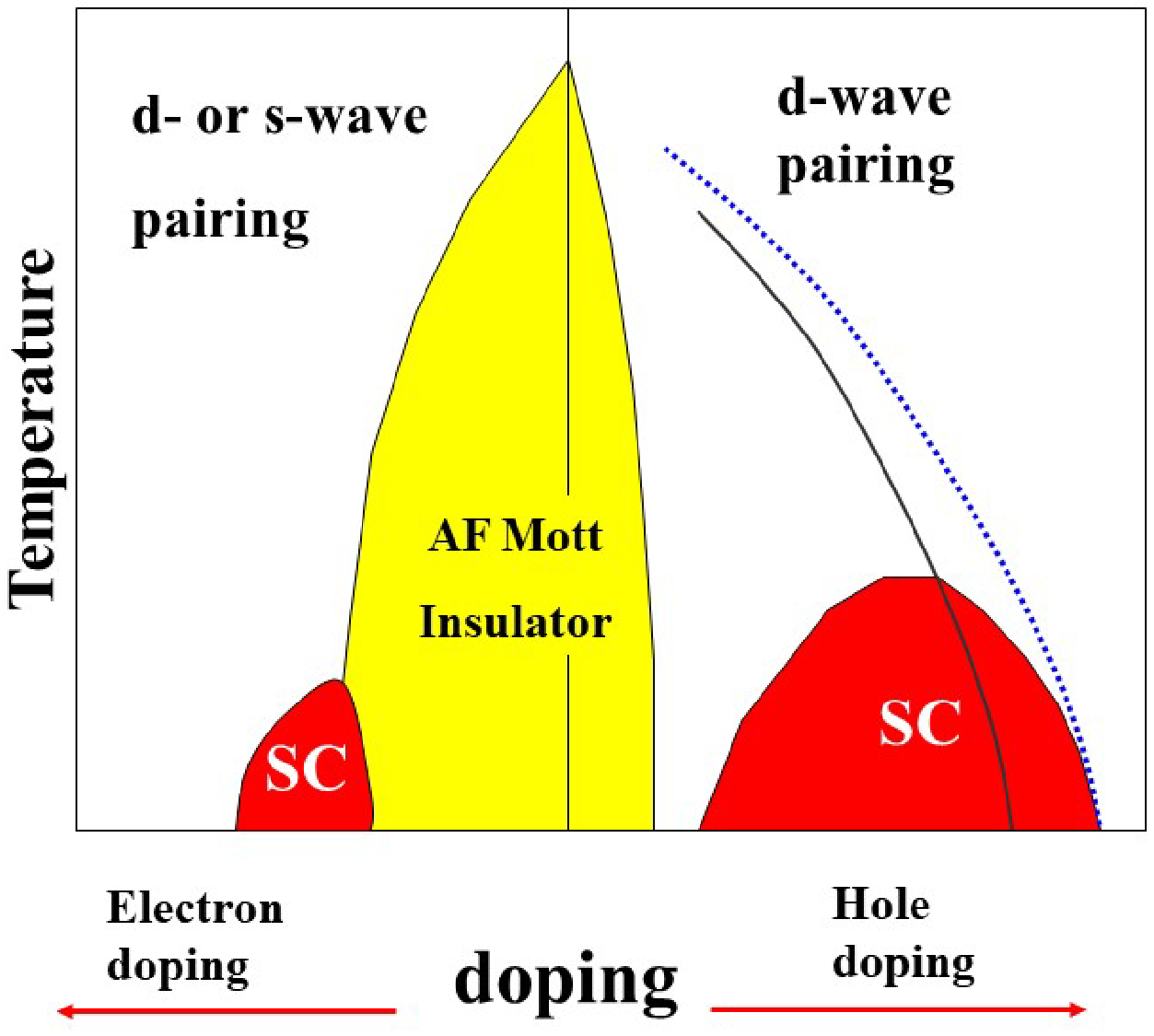}
\caption{A general electronic phase diagram of the
cuprate superconductors. The central yellow region shows the parent phase with a long range ordered antiferromagnetic insulating phase. The two
dome like highlighted by red color show the regions for superconductivity. In the
electron doped side, the AF order may enter the superconducting
dome, showing a coexistence of superconductivity and AF order. The pairing symmetry has been proved to be d-wave in the hole
doped region, it may be non-monotonic d-wave or s-wave pairing in the electron doped
region. The solid and dashed lines represent the pseudogap temperatures according to different definitions or experimental measurements.} \label{f3}
\end{figure}

No sooner after the discovery by M\"{u}ller and Bednorz, it was
found that the major players for superconductivity in the cuprates are
the $CuO_2$ planes in the middle of the pyramid constructed by the
octahedra of Cu and oxygen atoms. This plane is sketched in Fig.2. If we
have a look at the formulae of some parent phase of cuprate systems, such as $La_2CuO_4$, a natural counting on the ionic
state would lead to $Cu^{2+}$ with 3$d^9$, and $O^{2-}$ with 2$p^6$
as the outermost shell of electron orbitals. Taking the structural
parameters of the system into consideration, a simple
density-functional-theory (DFT) calculation will find out that the
material should be a metal with $Cu-3d$ orbitals as the conduction band, forming a large electron like Fermi surface surrounding the $\Gamma$ point, while
later experiments show that it is an insulator due to Mottness, thus it is called as a
Mott insulator\cite{AndersonMott}. Many transport properties
indicate that the resistivity close to the undoped case cannot be
described by a band insulating picture, rather it is described by a law
$\rho(T)\propto log(1/T)$ which manifests itself some exotic
reasons. Almost simultaneously, researchers found that the undoped phase,
here called as the parent phase of the cuprate superconductors, has
a long range anti-ferromagnetic order with an ordering temperature
at around 300 K and the super-exchange energy $J \approx 150 meV$. Clearly the insulating behavior of the parent phase is due to the strong correlation effect. A Hubbard model based picture, namely the Zhang-Rice singlet\cite{ZhangRice} model, was proposed to interpret the electric conduction in the doped cuprates. In the cuprate system, due to its crystalline field effect, the $3d_{x^2-y^2}$ orbital is lifted to the highest energy and is half filled. However, because of the strong Hubbard U (6-8 eV), this half-filled band will be split further into lower and upper Hubbard bands with a large U between them. In this case, the oxygen $2p$ band will hybridize with the lower Hubbard band, thus the doped holes would accommodate in this Cu-3d and O-$2p$ hybridized band. The Zhang-Rice singlet model postulates that each doped hole on the oxygen site will couple to the $Cu^{2+}$ cation and form a spin singlet state, it is this composed singlet that moves, leading to electric conduction. Meanwhile there are many other models to interpret this interplay of charge and spin in the doped cuprate, for example the t-J model incorporating a SU(2) symmetry\cite{WenLee}, and the phase string model\cite{WengZY}, etc. All these different models can also interpret the finite coherent DOS at the Fermi energy and a superconducting state with a diluted superfluid density arising from a strongly correlated background.
A typical electronic phase diagram is shown in Fig.\ref{f3}. More interestingly,
this insulating feature will leave the way to a strange metallic
behavior if we dope it with holes (hole doping) or electrons
(electron doping). Above a certain threshold of doping ($p \approx
0.05$ for example in $La_{2-x}Sr_xCuO_4$) the superconductivity emerges at a finite temperature. The $T_c$ versus doping normally
exhibits a dome like shape with the optimal $T_c$ around a doping
level of about $p = 0.16$ in the hole doped side. In the overdoped
side the $T_c$ drops down again and superconductivity vanishes at
about $p \approx 0.25-0.30$. Phase separation picture was proposed
to explain the dropping down of the superfluid density and the
superconducting transition temperature in the overdoped
region\cite{WenHHOverDope}. Recently, mutual inductance measurements on the systematically doped thin films illustrate also a rough linear correlation between the superfluid density and $T_c$ in the overdoped region, manifesting also an non-BCS mechanism\cite{BozovicNature}.

The normal state of the cuprate superconductors is also
drastically deviating from the very base for building up the BCS
paradigm, the Landau-Fermi liquid picture. One of the exotic
features here is the appearance of the so-called pseudogap far above
$T_c$, as shown in Fig.\ref{f3} by the solid and dashed lines. According to the different
definitions, the pseudogap temperature $T^*$ is also quite
different. For example, the Knight shift in the NMR measurements
shows a decrease at a quite high temperature. This has been
regarded as the opening of a spin gap\cite{Alloul}. The pseudogap
can also be detected in the temperature dependence of resistivity in
many cuprate systems. Below the pseudogap temperature, the
resistivity shows a linear behavior with a kink at the pseudogap
temperature. Similar behavior occurs in the specific heat data which
exhibits also a decrease of the electronic specific heat coefficient below the
pseudogap temperature\cite{Loram}. While the Nernst effect measurements
reveal a strong signal which may correspond to the existence of the
simultaneous vortex-antivortex excitations\cite{XuZA}. The more direct evidence
for the pseudogap is coming from the angle resolved photoemission
spectroscopy (ARPES)\cite{ShenZX}. One appealing picture for the
normal state is the gapped feature near the anti-nodal area due to
the correlation effect, while the electrons near the nodal point
have still finite life-time recovering a certain weight of
quasiparticles. This leads to the so-called truncated ``Fermi surface" or Fermi arc. Although to reckon on the number of the
conduction electrons through the well known Luttinger theorem is
very questionable, while the superconducting condensation in the
cuprates seem to be achieved by building up a gap on the so-called
Fermi arc near the nodal point. When temperature is increased above
$T_c$, the gap on the Fermi arc vanishes and the ``Fermi arc metal"
is recovered as the normal state\cite{WenLee}. Connected with this pseudogap feature, there are many other ``intertwined orders/phases" existing in this region, such as the charge density way (CDW) order, pair-density wave (PDW) order, spin-glass, etc. A recent overview gives a nice summary about this complexity\cite{KeimerNature}. Thus we see many very exotic
features of the normal state of the cuprate superconductors. These
abnormal features in the normal state of cuprate superconductors are
certainly shaking the very base for the BCS picture. It is still
under a hot debate how these facts can be put together to reach a
reconciled picture. Although some people suggest that the BCS
picture may be still applicable to interpret the superconductivity in
the cuprate system, while significant modifications have to be
undertaken, especially the focus is about whether the retarded weak
coupling picture, as the case for phonon mediated pairing, is still
applicable here.

The discovery of iron based superconductors\cite{Hosono} seems to help
unraveling the puzzle of unconventional superconductivity. Up to date, many FeAs or FeSe
based superconductors have been synthesized, which enriches the families
of the iron based superconducting systems\cite{WenHHReview}. A common
knowledge now about the iron based superconductors is that the major electric conduction and the
superconductivity are fulfilled by the FeAs or FeSe based planes.
Theoretical calculations indicate that the Fe has a cationic state
of $Fe^{2+}$ with 6 electrons in the five 3d orbitals, namely
$d_{xy}$, $d_{xz}$, $d_{yz}$, $d_{x^2-y^2}$ and $d_{3z^2-r^2}$.
One can estimate the crystalline field splitting of
these five 3d orbitals in the system, which gives a quite weak
effect. Therefore in principle, all the five 3d orbitals cross the
Fermi energy leading to several sets of Fermi surfaces, two
hole-like cylindrical Fermi pockets and a small 3D like electron
pocket in the Brillouin zone center ($\Gamma$ point), and two
more electron Fermi pockets in the zone corner ($M$ point).
For most FeAs based superconductors, the Fermi surface contains a set of hole pockets centering at $\Gamma$ and a set of electron pockets near M point, as sketched in Fig.4(b). Early but swift theoretical calculations indicate that the phonon
spectrum and the electron-phonon interaction does not support to
give superconductivity above 1 K\cite{Boeri}. Interestingly the
parent phase, for example, RFeAsO (R= rare earth elements) and
BaFe$_2$As$_2$ all show an antiferromagnetic order at temperatures
between 100-200 K. The wave vector of these AF phase measures
exactly the momentum displacement between the electron pockets
located at M($\pm\pi,0$) or ($0,\pm\pi$) to $\Gamma$(0,0) in most
families of iron based superconductors. There are some preliminary
evidence to show that the iron based superconductors also have the
pseudogap effect\cite{Pseudogap,Shimojima}, although it is quite rare compared with that
in the cuprate superconductors. Meanwhile the normal state,
especially those in phosphorous doped systems (LaFePO, or
BaFe$_2$As$_{2-x}$P$_x$) indeed indicates a super-linear temperature
dependence of resistivity near the possible quantum critical point
where the AF vanishes\cite{MatsudaQCP}. Therefore we have the strong
reason to believe that the superconductivity in the iron based
superconductors should be closely related to the antiferromagnetic
phase\cite{Imai}, and may have the same origin as the cuprates.

Recently, superconductivity up to 15 K was found in hole doped $Nd_{1-x}Sr_xNiO_2$\cite{Ni112SC}. Interests arose immediately after this discovery since this infinite layer system may also have a $3d^9$ orbital as the dominant one for the electric conduction. The discovery of superconductivity has also been extended to other compounds, for example $Pr_{1-x}Sr_xNiO_2$ and $La_{1-x}Ca_xNiO_2$. The calculations based on DFT reveal that for the parent phase $Nd(La)NiO_2$, the major bands crossing Fermi energy are $Ni-3d_{x^2-y^2}$, $Nd-5d_{3z^2-r^2}$, and $Nd-5d_{xy}$ with some hybridizations. Thus it is clear that the major contribution of superfluid density comes from the orbital of $Ni-3d_{x^2-y^2}$, but the $Nd-5d$ bands also contribute partially to the DOS at the Fermi level\cite{Norman,Thomalle}. A common understanding is that the $Ni-3d_{x^2-y^2}$ band constructs a large hole pocket around $\Gamma (0,0,0)$ at the $k_z = 0$, which changes into an electron pocket at the $k_z$ = $\pi$ due to the dispersion along $k_z$ direction. The hole pocket at $k_z$ = 0 looks like that of underdoped cuprate; while on the cutting plane at $k_z$ = $\pi$, the electron pocket shrinks and becomes similar to the Fermi pocket in overdoped cuprate. There are two more 3D electron pockets, one is surrounding $\Gamma$ which is contributed by the hybridization of the $Nd-5d_{3z^2-r^2}$ and $Ni-3d_{3z^2-r^2}$, another one surrounding A($\pi,\pi,\pi$) is constructed by the hybridization of $Nd-5d_{xy}$ and $Ni-3d_{3z^2-r^2}$. A spin excitation spectrum has been measured in the undoped system $NdNiO_2$, which indicates a wide band spin excitations up to an energy of 200 meV\cite{LeeWS}. Based on the scheme of repulsive interactions\cite{Thomalle}, $d$-wave pairing as the dominant one has been predicted on the central hole pocket. By measuring scanning tunneling microscopy, two types of gaps are observed\cite{WenNi112}, one has a $d$-wave gap function with gap maximum about 3.9 meV, another one exhibits a full gap about 2.35 meV. Some time a mixture of the two types of gap features appears on one single spectrum, which indicates that the $Nd_{1-x}Sr_xNiO_2$ thin film is a multigap system and probably with a $d$-wave pairing tendency as the dominant one. Due to the difficulties of synthesizing good samples, more results are desired to unravel the mystery of superconductivity in the nickelate systems.

\begin{figure}
\vspace{-3.0em} %
\centering
\setlength{\abovecaptionskip}{0.0cm}
\includegraphics[width=12cm]{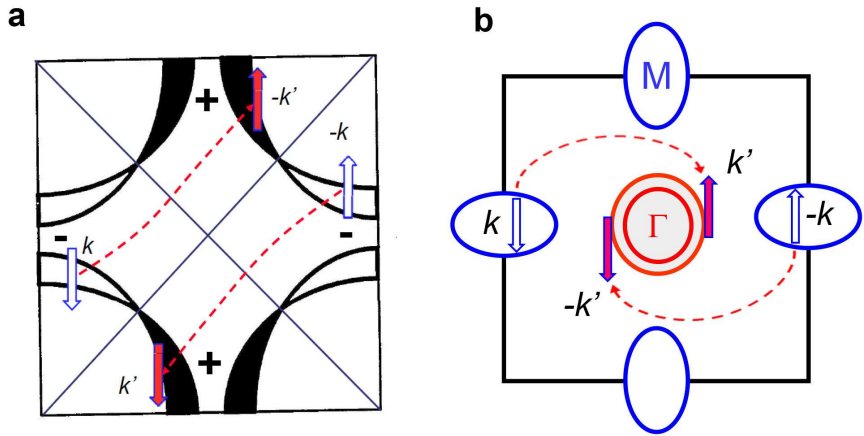}
\caption{Cartoon picture for the pair-scattering process based on a retarded weak
coupling in (a) cuprates and (b) iron based superconductors. The open arrows represent two electrons with opposite momenta and spins before the pair-scattering; the filled arrows show the final states after the pair-scattering. The dashed lines show the pairing channel in the pair-scattering process through a retarded interaction. If the interaction V is positive, this kind of pairing will induce sign change of the superconducting gaps in the momentum space, for example d-wave in cuprates and s$^\pm$-wave in iron-based superconductors.} \label{f4}
\end{figure}

\section{Preliminary understanding of the unconventional superconductivity}

If there exist strong Coulomb interactions $U$ among electrons, or the ratio between the repulsive coulomb interaction
and the band width $t$, i.e., $U/t$ is large, the free electron gas
or the so-called Landau-Fermi liquid picture immediately
becomes questionable. In this case one cannot assume a negligible potential energy in the electronic Hamiltonian, the latter has been adopted in building up the BCS picture. With the presence of strong correlations, the Hamiltonian has been considered
based on the $t-U$, or $t-J$ or $t-J-U$ models. Since tremendous
number of electrons need to be considered in the system, this multibody problem is not solvable. However, this problem may
be at least inspected by assuming a low energy approach. In this
case, a microscopic Hubbard-type Hamiltonian with a four fermion
interaction can still be written as\cite{Chubukov}

\begin{equation}
H=\sum_{k,\alpha}\varepsilon_k\psi^+_{k\alpha}\psi_{k,\alpha}+\sum_{k_i,\alpha_i}U^{\alpha_1,\alpha_2,\alpha_3,\alpha_4}_{k_1,k_2,k_3,k_4}\psi^+_{k_1,\alpha_1}\psi^+_{k_2,\alpha_2}\psi_{k_3,\alpha_3}\psi_{k_4,\alpha_4}.
\label{e6}
\end{equation}

Here $U^{\alpha_1,\alpha_2,\alpha_3,\alpha_4}_{k_1,k_2,k_3,k_4}$ represents the four electron interactions, $\psi^+_{k_i,\alpha_i}$ ($\psi_{k_i,\alpha_i}$) is the creation (annihilation) operator for electrons with the spin $\alpha_i$ and momentum $k_i$ (i=1,2,3,4). For a single band Hubbard model with local Coulomb interactions, we have

\begin{equation}
U^{\alpha_1,\alpha_2,\alpha_3,\alpha_4}_{k_1,k_2,k_3,k_4} = U\delta_{k_1+k_2-k_3-k_4}(\delta_{\alpha_1\alpha_4}\delta_{\alpha_2\alpha_3}-\delta_{\alpha_1\alpha_3}\delta_{\alpha_2\alpha_4}).
\label{e7}
\end{equation}

If only the interactions in the low energy area are considered, the
situation can be treated by using the so-called random phase
approximation (RPA) method\cite{ScalapinoRPA}. For a system with
very good Fermi surface nesting, the electronic system will be
instable to some long range ordered state, such as the
charge-density-wave (CDW) or the spin-density-wave (SDW) state. In
the iron based systems, the perfect nesting between the hole and
electron pockets may just satisfy this situation leading to the
formation of a long range SDW order in the parent phase in the FeAs based families. There are some arguments that the AF order is induced by the local super-exchange effect. By chemical doping, or applying a pressure, the perfect
nesting situation deteriorates, and the long range ordered SDW state
will be replaced by the state with strong antiferromagnetic spin
fluctuations (AF-SF). The elementary excitations of this state are
the so-called paramagnons. When an itinerant electron is passing
through a position, it will polarize the spin cloud around it,
this polarized spin cloud will be sensed by the second itinerant
electron passing by. This process will build up a pairing
interaction between the two electrons with a pairing interaction as

\begin{equation}
\Gamma_s(k,k^\prime)=\frac{3}{2}U^2\frac{\chi_0(q)}{1-U\chi_0(q)},
\label{e8}
\end{equation}

where $U$ is the repulsive energy, $\chi_0$ is the bare spin
susceptibility in absence of interactions, the subscript $s$ here
represents the singlet pairing channel. One can see that $\Gamma_s$
is also peaked at the wave vector $q$ but is positive. If we use
this pairing interaction and follow the BCS approach, the gap
function can be written as

\begin{equation}
\Delta_k=-\sum_{k^\prime}\Gamma_s(k,k^\prime)\frac{\Delta_k^\prime}{2E_k^\prime}tanh\frac{E_k^\prime}{2T}
\label{e9}
\end{equation}

Here $E_{k^\prime}=\pm\sqrt{\varepsilon_{k^\prime}^2+\Delta_{k^\prime}^2}$. In iron based superconductors, this interesting pairing picture was
first proposed by Mazin et al.\cite{Mazin}, and later further
formalized by several other groups using different theoretical
approaches\cite{Kuroki,WangZD,Hirschfeld1,LeeDH}. More details about
the pairing order parameter can be found in a recent
review\cite{Hirschfeld2}. The basic idea of weak coupling picture is illustrated in
Fig.\ref{f4}(a) and (b) for cuprates and iron based superconductors, respectively. It still remains unclear whether this weak coupling based picture applies to the cuprates, since the pairing is very strong and Fermi surface is incomplete (truncated). In iron based superconductors, especially in FeAs based superconductors, the two electrons on the electron (hole) FSs
(marked with the thick arrows indicating opposite directions of
spins) are scattered to the hole (electron) FSs by exchanging the
antiferromagnetic spin fluctuations (AF-SFs). If the pairing is established through exchanging the
AF-SFs, according to above equations, the so-called $S^\pm$ pairing
model is expected, i.e., the DOS on individual FS are fully gapped
(although the gap magnitude has some momentum dependent variation), but
the signs of the superconducting gaps on the electron and hole pockets should be
opposite. This kind of s-wave gap has been supported by the scanning
tunneling experiment already\cite{Hanaguri}. The recent experiment in observing the
in-gap quasi-particle states by non-magnetic Cu impurities in
$NaFe_{0.97-x}Co_{0.03}Cu_xAs$ may provide a more decisive evidence
for the $S^\pm$ pairing\cite{CuImpurity}. Furthermore, this sign reversal of gaps on the electron and hole pockets in $FeTe_{1-x}Se_x$ has been well checked by the phase referenced quasiparticle interference technique\cite{ChenMYPRB}. This RPA based picture may
also be extended to the case of cuprates for understanding the well
known $d$-wave pairing symmetry. As shown in Fig.\ref{f4}(a), since
there is only one set of Fermi surface in the cuprate, the
pair-scattering would lead to a sign change on the different
segments of the Fermi surface, this naturally leads to a $d$-wave gap
function.

\begin{figure}
\vspace{-3.0em} %
\centering
\setlength{\abovecaptionskip}{0.0cm}
\includegraphics[width=12cm]{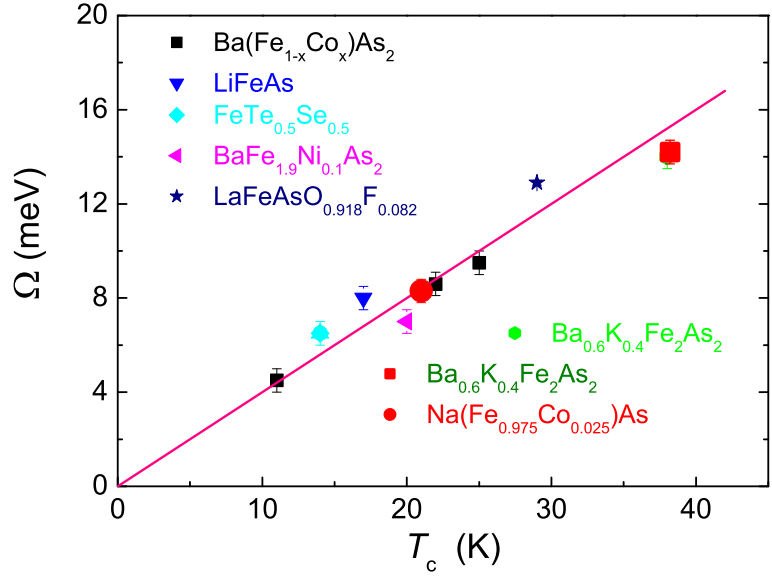}
\caption{ A roughly linear relation between the neutron resonance or bosonic mode energy and the
superconducting transition temperature $T_c$ in many iron based superconductors. The small sized filled symbols are data collected in the inelastic neutron scattering measurements, the two red bigger square and circle symbols are determined from the bosonic mode on the tunneling spectrum. This figure is made by quoting to the data in Ref.[52]} \label{f5}
\end{figure}

One of the consequences of the electronic system with the pairing gap
with opposite signs is the appearance of a resonance peak in the
imaginary part of the spin susceptibility at the wave vector of the
SDW. This resonance peak with a typical energy $\Omega\leq 2
\Delta_s$ can be detected with inelastic neutron scattering
experiment and has been observed in cuprates\cite{RPCuprate}, heavy
Fermion\cite{RPHeavyFermion} and recently in the iron pnictide
superconductors\cite{RPIron}. This resonance peak may be understood
as due to the spin flipping scattering when exciting a quasiparticle
from the $k$ to $k+q$ with the pairing gap with the opposite signs,
since in this case the BCS coherence factor is given
by\cite{ScalapinoRPA,Hirschfeld2}

\begin{equation}
I_{res} \propto \frac{1}{2}[1-\frac{\Delta(k)\Delta(k+q)}{E(k)E(k+q)}].
\label{e10}
\end{equation}

Here $I_{res}$ gives the intensity of the resonance peak, $E(k)$ is
the dispersion of the Bogoliubov quasiparticles in the superconducting state
$E(k)=\pm\sqrt{\varepsilon_k^2+\Delta^2(k)}$. One can see that, if the superconducting gap has a sign reversal at momenta connected by vector $q$, that will produce a peak at this particular vector. The RPA calculations tell that this peak should locate at a energy of $\Omega \leq 2\Delta_s$ depending on the correlation strength. Although the resonance
peak is regarded as the consequence of the sign-reversal gaps,
sometimes it is called as the ``egg" of the superconducting pairing
state, not the pairing glue for superconductivity, we must
emphasize that this ``egg" can be naturally traced back to the
scenario of the magnetic pairing, namely the
origin of pairing. Therefore, although the resonance peak may be due to the
spin-1 exciton mode of the sign reversal gaps, we can conclude that
it is only possible that the pairing is induced through the
electronic origin, either through exchanging the AF-SFs, or the superexchange itself, but certainly not through some non-magnetic origins, like exchanging phonons. In Fig.\ref{f5}, we present a collection of many resonant energies versus the superconducting transition temperature $T_c$ in different iron based superconductors. Here the filled symbols are data collected in the inelastic neutron scattering measurements, the two open symbols are determined from the bosonic mode on the tunneling spectrum. In some iron based superconductors, the
scanning tunneling spectroscopy measurements reveal extra peaks (humps)
outside the superconducting coherence peaks, which is explained as
the coupling of the quasiparticles with the spin resonance mode. In Fig.\ref{f5}, a simple
ratio $\Omega \approx (4.3\pm 0.5) k_B T_c$ is shown\cite{NatPhysWangZY}. This strongly suggests that the
neutron resonance is indeed closely related to
superconductivity. Similar linear relation between $\Omega$ and $T_c$ was also found in cuprate superconductors\cite{PNAS}. A study on the superconducting condensation energy measured through
specific heat and the intensity of the resonance peak in the $Pr_{0.88}LaCe_{0.12}CuO_{4-\delta}$ finds that both quantities decay with the applied magnetic field in the same
way\cite{PNAS}, indicating a close relationship between the resonance mode energy and superconductivity.

\begin{figure}
\vspace{-3em} %
\centering
\setlength{\abovecaptionskip}{0.0cm}
\includegraphics[width=10cm]{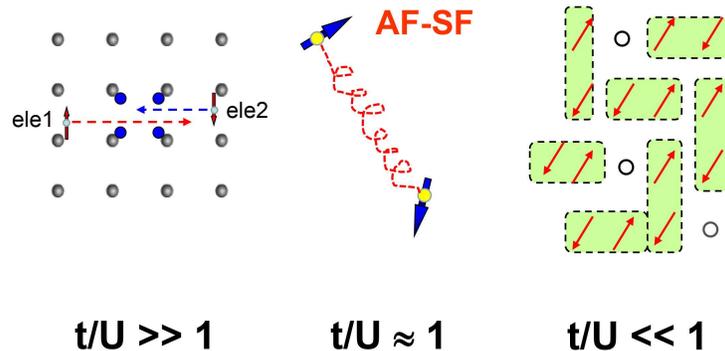}
\caption{Evolution of the pairing from weak coupling to strong coupling, depending on the ratio $U/t$. The left hand picture shows the case for $U/t<<1$, the BCS scheme works well. Here the four blue circles represent the distortion of the atoms due to the electron-phonon interaction. This distortion will be ``seen" by the second electron passing by in a retarded duration. The right hand one represents the possible situation of strong correlation effect where the local
super-exchange may dominate the pairing. In the middle, the $U/t\approx 1$ which may be applicable for the case of AF-SF mediated pairing.} \label{f6}
\end{figure}

\begin{figure}
\centering
\setlength{\abovecaptionskip}{-1.cm}
\includegraphics[width=12cm]{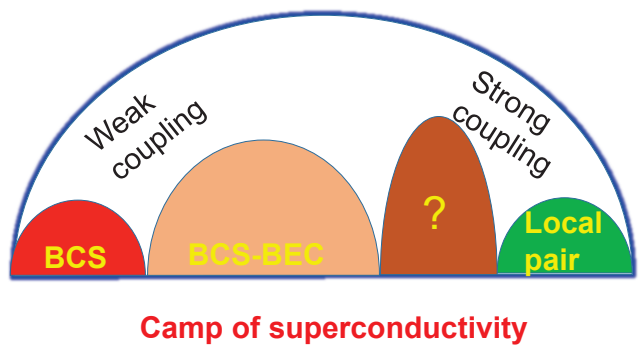}
\caption{Schematic camp for superconductivity. The BCS theory with weak coupling occupies a small region in the left hand side. We need a general or global camp to cover all possibilities for superconductivity, from weak coupling to strong local pairing.} \label{f7}
\end{figure}

In above discussion, we have shown that the low energy approach, or
called as the picture of weak coupling limit can give a reasonably good
understanding of the superconductivity in many unconventional
superconducting systems. In the systems with moderate Coulomb
repulsion energy $U$, this seems to work quite well. In some systems
with extremely large ratio $U/t$, it is questioned whether this
picture based on retarded scattering can still be valid. In this
case, some people argue that the local super-exchange $J$ dominates
the pairing process, leading to the simultaneous
pairing\cite{LocalPair1,LocalPair2} with a very strong pairing
strength. One of the extreme cases is the so-called
resonance-valence-bond (RVB) spin liquid state for the $s = 1/2$
systems, for example in the cuprates. In this case, the spin singlet
pairs are naturally formed yielding a highly fluctuating quantum RVB
state.\cite{RVB} This is certainly a very interesting picture which
may get an indirect support from the strong gap, very strong superconducting
fluctuation, or residual Copper pairs in the normal state far above
$T_c$. The convincing evidence for this RVB state is still lacking.
In Fig.\ref{f6} we present a schematic show for the superconducting
pairing for the cases ranging from the weak coupling (e.g., phonon mediated)
pairing to the extreme case of the simultaneous pairing through the
local super-exchange. The situation evolves with the ratio of $t/U$: (1) $t/U\gg1$, weak coupling induced pairing; (2) $t/U \approx 1$, intermediate correlation induced pairing; (3) $t/U\ll1$ strong correlation and local pairing could occur. This picture follows a similar thought as
that when considering the subtle balance between the correlation
effect and the effective Drude weight for optimized
superconductivity in a variety of compounds\cite{Basov}.
In the middle of Fig.\ref{f6}, we have a delicate balance between
the two extreme cases: complete itinerant ($U/t<<1$) and strong
correlation ($U/t>>1$). The middle case may apply to the iron based superconductors, which still needs the
retarded interaction as the pairing origin, but through exchanging
the paramagnon or AF-SFs, not phonons. At the moment or in the
near future, debate will go on about whether we need to have a more exotic
picture which assumes the pairing completely coming from the local
strong superexchange. Based on this idea, in Fig.\ref{f7}, we sketch a big camp for superconductivity. The BCS theory occupies only a small area in the weak coupling limit. It remains to know how the pairing can evolve smoothly from the weak coupling limit to the extremely strong coupling limit. It is expectable that the condensation process for superconductivity will also evolve from the BCS type to BEC type, the superconducting transition temperature is determined by the gap in the BCS case, but by the phase stiffness in the BEC case\cite{Kievelson,Randeria}.

\section{Some empirical rules for the exploration of high temperature superconductors}
It is always enthusiastic to explore new high temperature superconductors. However, beside those of pressurized hydrides, no theory can really predict how to get success, thus it has been a quite frustrated task. Even so, some empirical rules may be summarized here. Here we would like to propose several key points which may be the necessary ingredients for exploring high temperature superconductivity at ambient pressure.

\subsection {Subtle balance between Mottness and itinerancy of $d$ orbital electrons}
As illustrated by Qazilbash et al.\cite{Basov}, the unconventional superconductors should have a delicate balance between Mottness and itineracy which governs the effective DOS at the Fermi energy. Strong Mottness may manifest a large superexchange energy $J$ for the local pairing, but the ground state of that may be an insulator. Thus we need to effectively separate the dual roles of the $d$-orbital electrons. The 3$d$ orbital electrons possess by them-self a possibility to balance between these two factors. Thus it may give more chances in the compounds with the 3$d$ transition metal elements, like Cr, Mn, Fe, Co, Ni, Cu. Some 4$d$ elements may also be considered, such as Ru, Pd, Rh etc.

\subsection{Bad metal in vicinity of an AF order}
 Recalling both the iron based and cuprate superconductors, it is known that the parent phases are either Mott insulators or bad metals with or near an AF order. Most importantly, this AF order can be easily suppressed by chemical doping or applying a pressure. To satisfy this condition, it is important to have ions with lower magnetic moments, for example below 2$\mu_B$. The spin density wave order due to the Fermi surface instability would help. To avoid the large magnetic moments, one may request the low filling or even number filling states of the 3$d$ or 4$d$ orbitals, such as the Cu-3$d^9$ or Fe-3$d^6$ cationic states.  This is actually the case in iron based superconductors. In BaFe$_2$As$_2$, for example, the cationic state is Fe$^{2+}$, which yields a magnetic moment generally smaller than 2$\mu_B$. In the cuprate, for example, each Cu ion in $La_2CuO_4$ has a magnetic moment of 0.5-1.0 $\mu_B$. The strong quantum fluctuation of the gauge field near the quantum critical point (QCP) may also help. An odd filling number of 3, 5, 7 of the $d$ orbital will induce a strong Hund's coupling leading to strong magnetic moments and localization of electrons.

\subsection{Shallow and flat band near the Fermi energy}
Some shallow and flat bands, or saddle points near the Fermi energy will result in the von Hove singularity of DOS, this is certainly helpful for the pairing instability. In iron based superconductors, it is known that some bands are very shallow, which yields a high DOS. In addition, a shallow band possesses a small Fermi energy $E_F$, which leads to the small ratio $E_F/\Delta_s$. This quantity measures the Cooper pair numbers within a coherence length $\xi$ or volume $\xi^3$. If this value is small, one may expect a BEC type superconducting transition. Therefore, to have both strong pairing and high superfluid density, we probably need a system with both shallow and wide bands and together with a strong interband scattering, the shallow band produces a strong local pairing, the wide band contributes a high superfluid density.

\subsection{Anion mediated superexchange may help}
In cuprates and iron based superconductors, there are anions to mediate the superexchange for the antiferromagnetic order or spin fluctuations. Thus in the dominant conductive planes, some atomic chains like T-A-T are necessary, where $T$ represents a transition metal with $d$ orbital electrons, $A$ is an anion. For example, in cuprates the pairing through $J_1$ superexchange is established by the Cu-O-Cu bonds. While in iron based superconductors, the pairing may be established by the $J_2$ superexchange which is generated by the Fe-As-Fe bonds. However, this requirement may not be that strict for high $T_c$ superconductivity. In many binary alloy compounds, unconventional superconductivity coexists with the magnetic excitations. Thus for unconventional superconductivity, the anion bridge may not be the necessary component, but may be helpful for stabilizing the structure.

\section*{Acknowledgement}
We acknowledge the useful discussions with Qianghua Wang, Ilya Eremin, Peter Hirschfeld, Andrey Chubukov, Igor Mazin. This work was supported by
NSF of China with Grant Nos. 11927809, NSFC-DFG1206113001(ER463/14-1), and the Ministry of Science and Technology of China.

\end{document}